\documentclass[layout=onecolumn, journal=jpca]{achemso}

\setkeys{acs}{chaptertitle = true}

\usepackage{geometry}
\usepackage{graphicx,subfigure,mciteplus}
\usepackage{amssymb, amsmath}
\usepackage[flushleft]{threeparttable}
\usepackage{epstopdf, booktabs}
\usepackage{multirow,bigdelim}
\usepackage{ragged2e}

\makeatletter
\g@addto@macro\TPT@defaults{\footnotesize}
\makeatother

\usepackage{graphicx}
\usepackage{dcolumn}
\usepackage{bm}
\usepackage{longtable}
\usepackage{multirow}
 
\SectionNumbersOn

\author{Jason M. Montgomery}
\affiliation{Department of Chemistry, Biochemistry, and Physics, Florida Southern College, Lakeland, FL 33801}
\author{David A. Mazziotti}
\email{damazz@uchicago.edu}
\affiliation{Department of Chemistry and The James Franck Institute, University of Chicago, Chicago, IL 60637}

\title[Maple QuantumChemistry]{Maple's Quantum Chemistry Package in the Chemistry Classroom}

\begin{document}

%
%
%
%
%

\begin{abstract}

An introduction to the Quantum Chemistry Package (QCP), implemented in the computer algebra system Maple, is presented. The QCP combines sophisticated electronic structure methods and Maple's easy-to-use graphical interface to enable computation and visualization of the electronic energies and properties of molecules. Here we describe how the QCP can be used in the chemistry classroom using lessons provided within the package. In particular, the calculation and visualization of molecular orbitals of hydrogen fluoride, the application of the particle in a box to conjugated dyes, the use of geometry optimization and normal mode analysis for hypochlorous acid, and the thermodynamics of combustion of methane are presented.

\end{abstract}


\section{Introduction}

\label{sec:introduction}

With the advent of high-performance computing, sophisticated electronic structure methods, and available quantum chemistry software packages, quantum calculations are now more powerful than ever. But to the undergraduate chemistry student, the abstractness of quantum chemistry and the technical nature of electronic structure methods can be dizzying.  Introductory concepts on the quantum mechanical nature of atoms and molecules are introduced in first year courses and continue to be relevant throughout the chemistry curriculum, but exposure to quantum calculations may be very limited, if not absent, in part due to a lack of access to affordable and easy-to-use electronic structure software.

Here we discuss the Maple Quantum Chemistry Package (QCP), also known as the Quantum Chemistry Toolbox, for use in the chemistry or physics classroom.  The QCP was specifically designed and developed for the computer algebra system Maple.\cite{MapleQCP,Maple}  Maple can be an invaluable tool in the chemistry classroom that combines a symbolic computation engine, efficient numerical algorithms, and visualization tools, all accessible through a user-friendly graphical interface.\cite{ed074p1491, ed300032f,ed061p629,ogilvie,gander} Accordingly, the QCP adds sophisticated quantum chemistry~\cite{griffiths,jensen,levine, sakurai,lowe,atkins, cramer,McQuarrie,mazziotti1,helgaker,mcquarrie2, szabo,cr2000493,pyscf} functionality to Maple to provide an extensive, easy-to-use platform for the computation of the electronic energies and properties of molecules.

Included in the QCP is a set of {\em Curricula} and {\em Lessons} that represent the ways in which the QCP can be integrated in the classroom.  Curricula are separated into General Chemistry, Physical Chemistry (both Quantum and Statistical Thermodynamics),  Advanced Physical Chemistry, and Physics.  Each curriculum contains links to corresponding lessons that may be imported as a Workbook and executed.  In general, each lesson provides an overview of learning objectives and already contains much of the required code and a narrative that guides a student through the lesson.

In what follows, we present portions of four lessons provided in the Quantum Chemistry Package (QCP) to give the reader an idea of how the QCP might be exploited in the chemistry classroom. The lessons are {\em Molecular Orbitals of Hydrogen Fluoride} (Sec.~\ref{sec:HFMOs}), {\em Particle in a Symmetric Box} (Sec.~\ref{sec:particle-box}), {\em Geometry Optimization and Normal Modes} (Sec.~\ref{sec:normal-modes}), and {\em Calculating Reaction Thermodynamics for Combustion of Methane} (Sec.~\ref{sec:thermo}).   It is important to note that each activity can imported as a worksheet and can be modified as necessary by students or instructor to fit the needs of the class or assignment.

\section{Curriculum Using the Quantum Chemistry Package}

\label{sec:curriculum}


\subsection{Initializing the Quantum Chemistry Package}

\label{sec:initialize}

Any application that utilizes the QCP begins with loading the package using the command {\em with(QuantumChemistry)}:  \\

\noindent \ \includegraphics[scale=0.37]{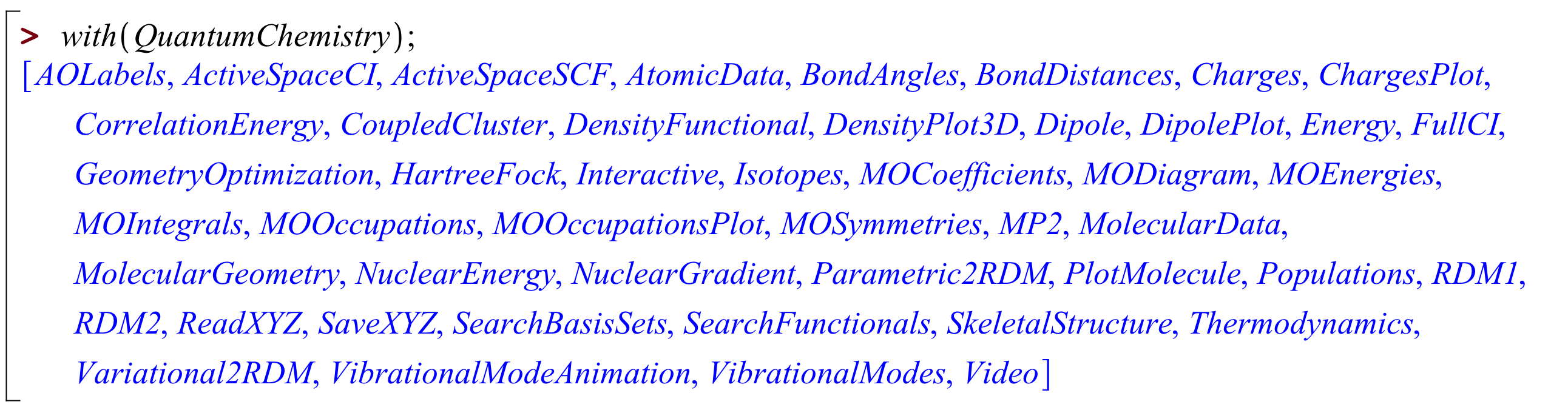}

\noindent Output of this command provides a list of the contents of the package.  Using Maple's {\em Search} toolbar, the user can seek help on any function included in the package, including the various options available for each function as well as examples.


\subsection{Molecular Orbitals of Hydrogen Fluoride}

\label{sec:HFMOs}

One activity provided with the QCP involves the calculation and visualization of molecular orbitals (MOs) for hydrogen fluoride (HF). The activity might be useful for an upper-level physical chemistry course, in which students can make connections between an underlying level of theory (electronic structure method and atomic orbital basis) and the resulting molecular orbital properties, or even organic chemistry or foundational general chemistry courses, in which students can make qualitative connections between atomic, molecular orbitals, and molecular properties.

After loading the QCP with the {\em with(QuantumChemistry)} command, the HF molecule is defined as a list of lists, with distance units being  angstroms by default:\\

\noindent \  \includegraphics[scale=0.4]{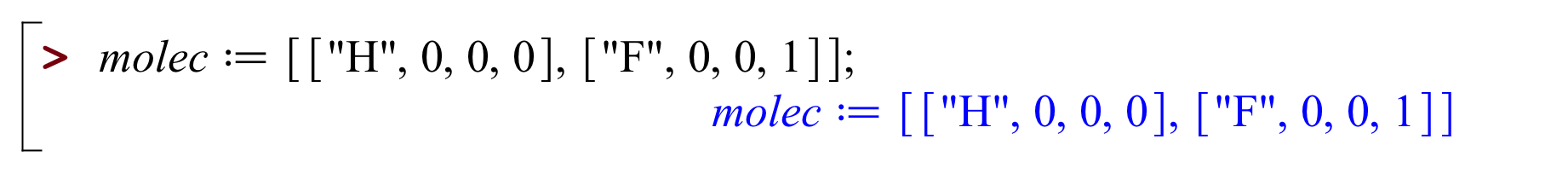}.

\noindent The user can use the {\em GeometryOptimization} function to find the minimum energy geometry:

\noindent \ \includegraphics[scale=0.44]{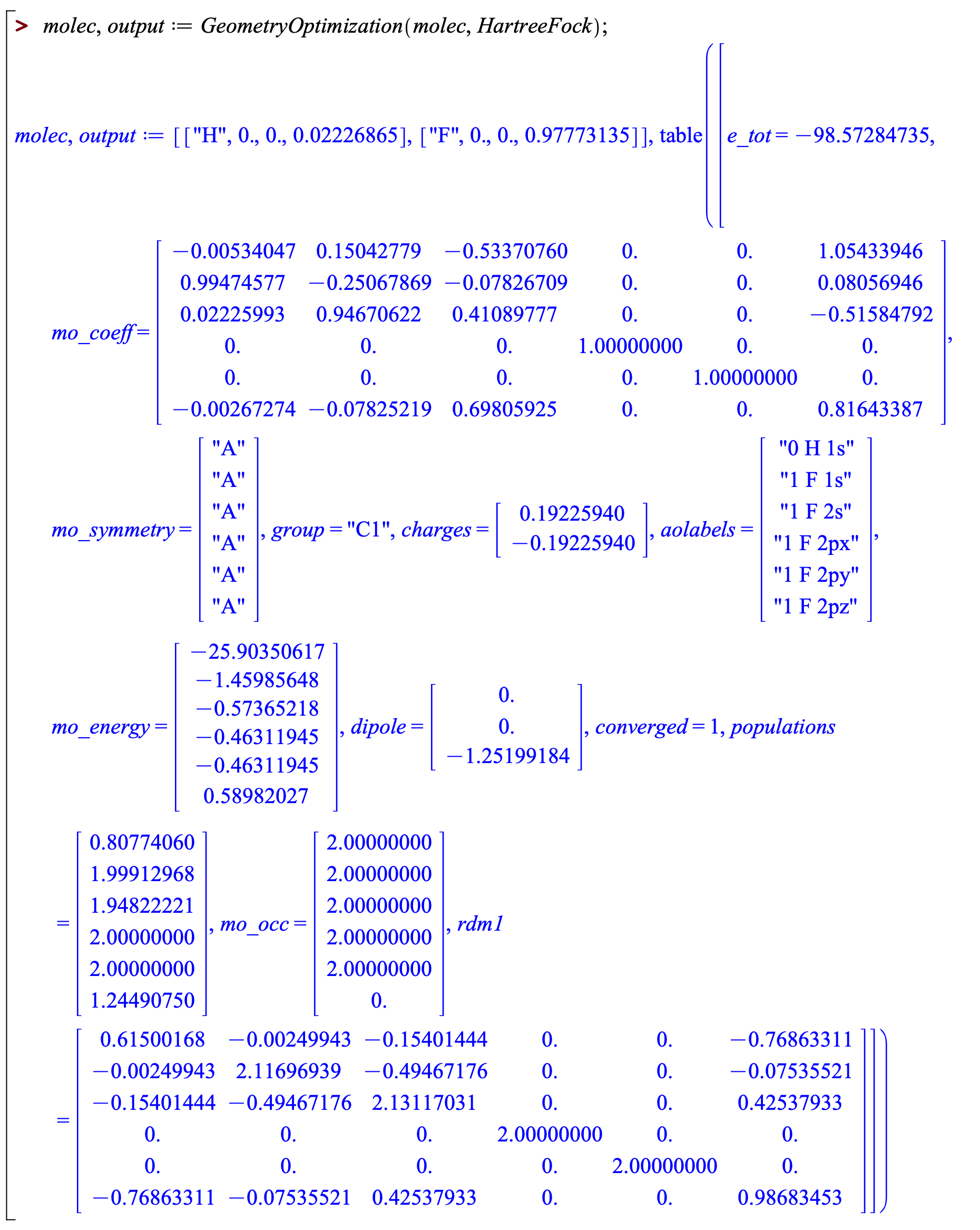} \\

\noindent In the input the user can specify through optional keyword arguments any method, basis, charge, spin, etc. In this case, we use Hartree-Fock method\cite{hartree,fock,roothan,szabo} with the default STO-3G basis set.  The output of the function provides an updated molecular geometry list as well as calculation results, including total energy,  molecular orbital energies,  molecular orbital occupation numbers,  molecular orbital coefficients, and even Mulliken charges on each atom.

A powerful feature of the QCP is the ability to perform {\em ab initio} calculations and visualize results with simple Maple commands.  For this activity, we visualize each of the five occupied MOs as well as the lowest unoccupied MO (LUMO) using the {\em DensityPlot3D} command:\\

\noindent \ \includegraphics[scale=0.5]{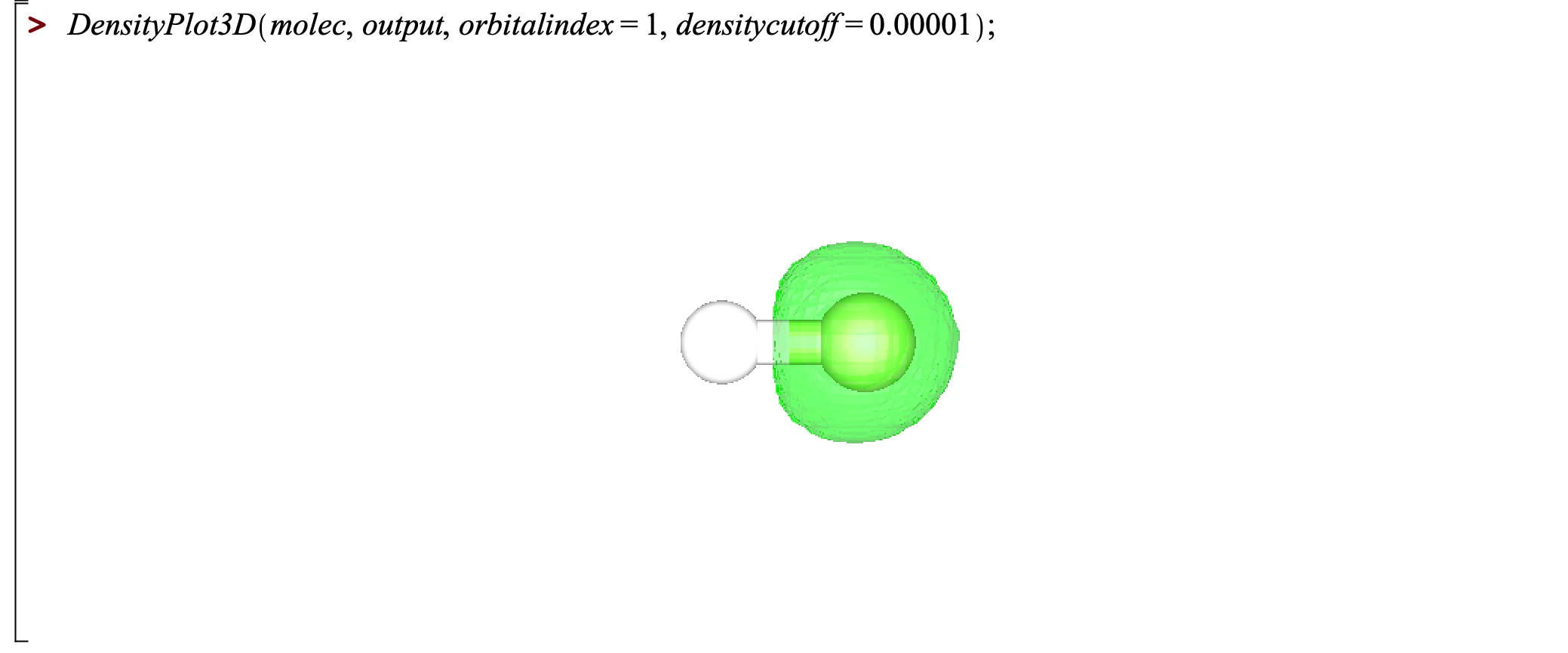}

\noindent \ \includegraphics[scale=0.5]{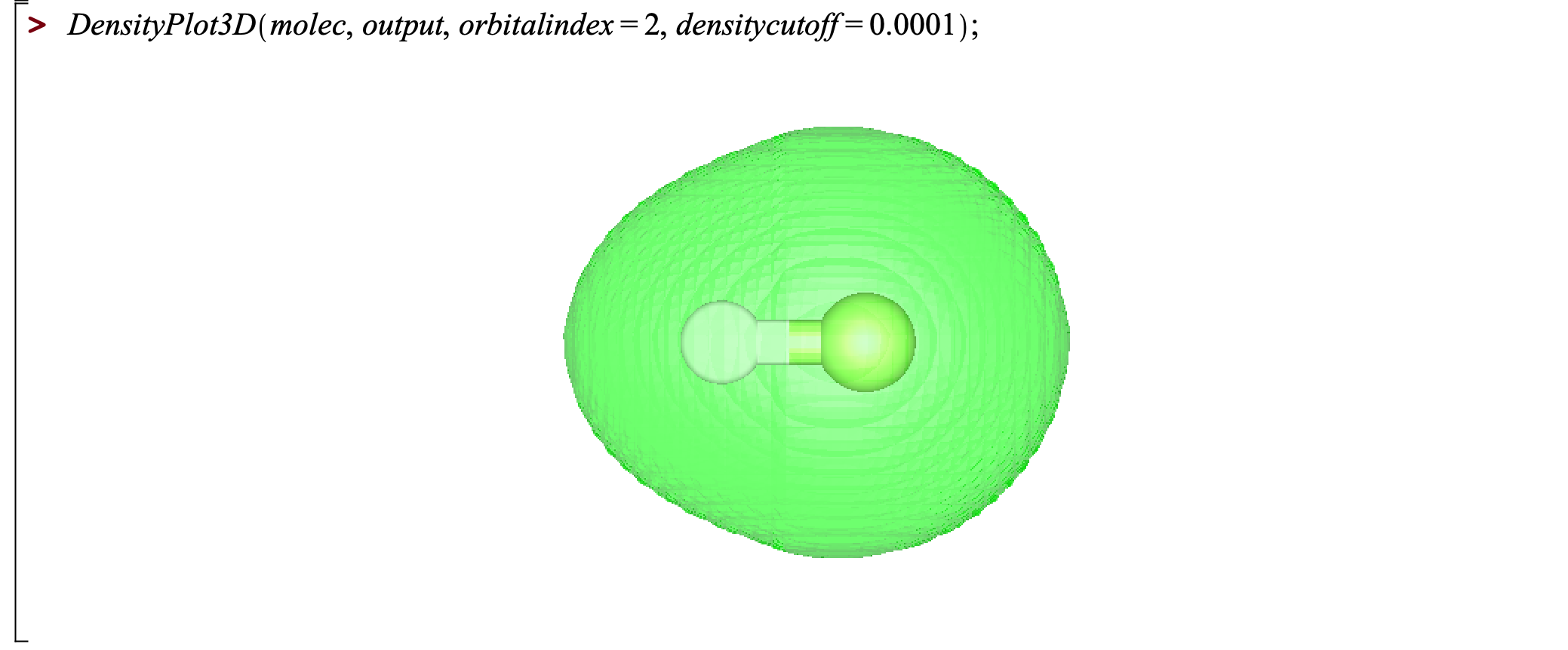}

\noindent \ \includegraphics[scale=0.5]{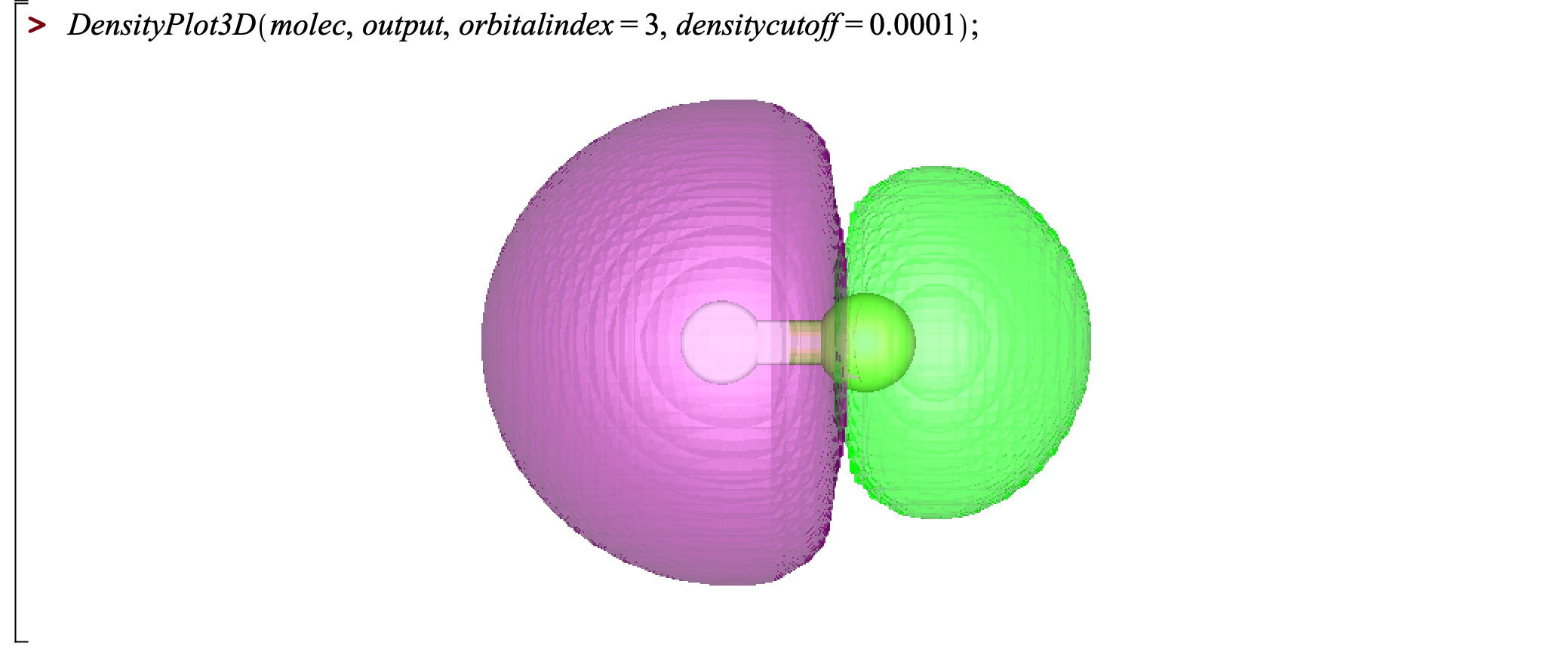}

 \noindent \ \includegraphics[scale=0.5]{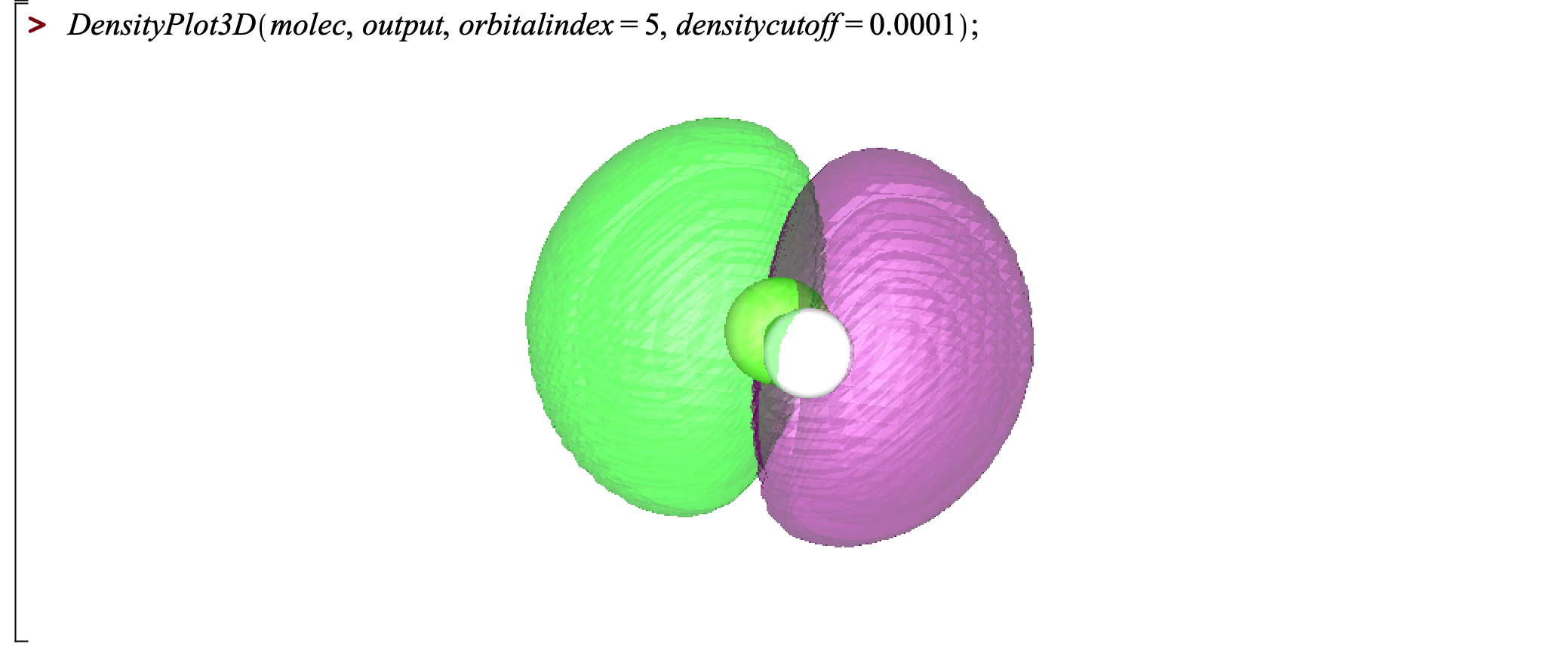}

\noindent \ \includegraphics[scale=0.5]{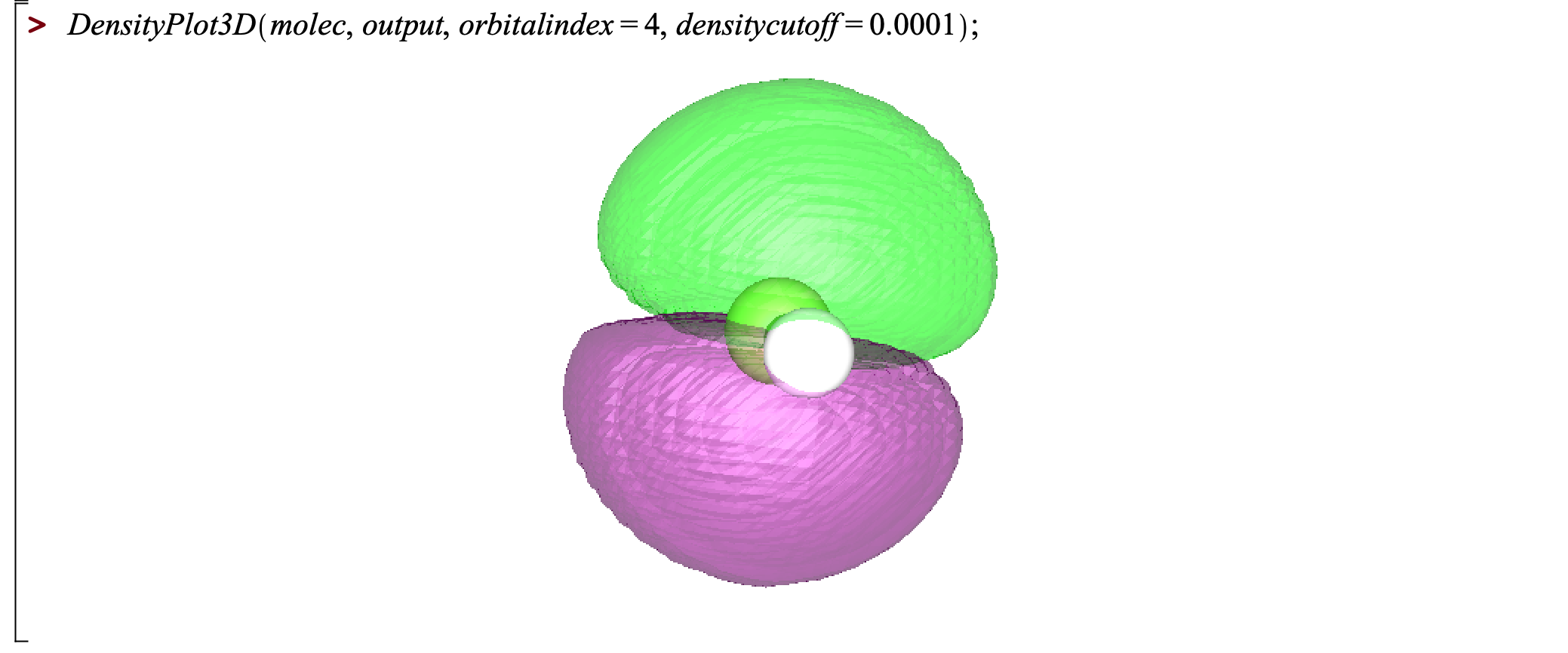}

\noindent \ \includegraphics[scale=0.5]{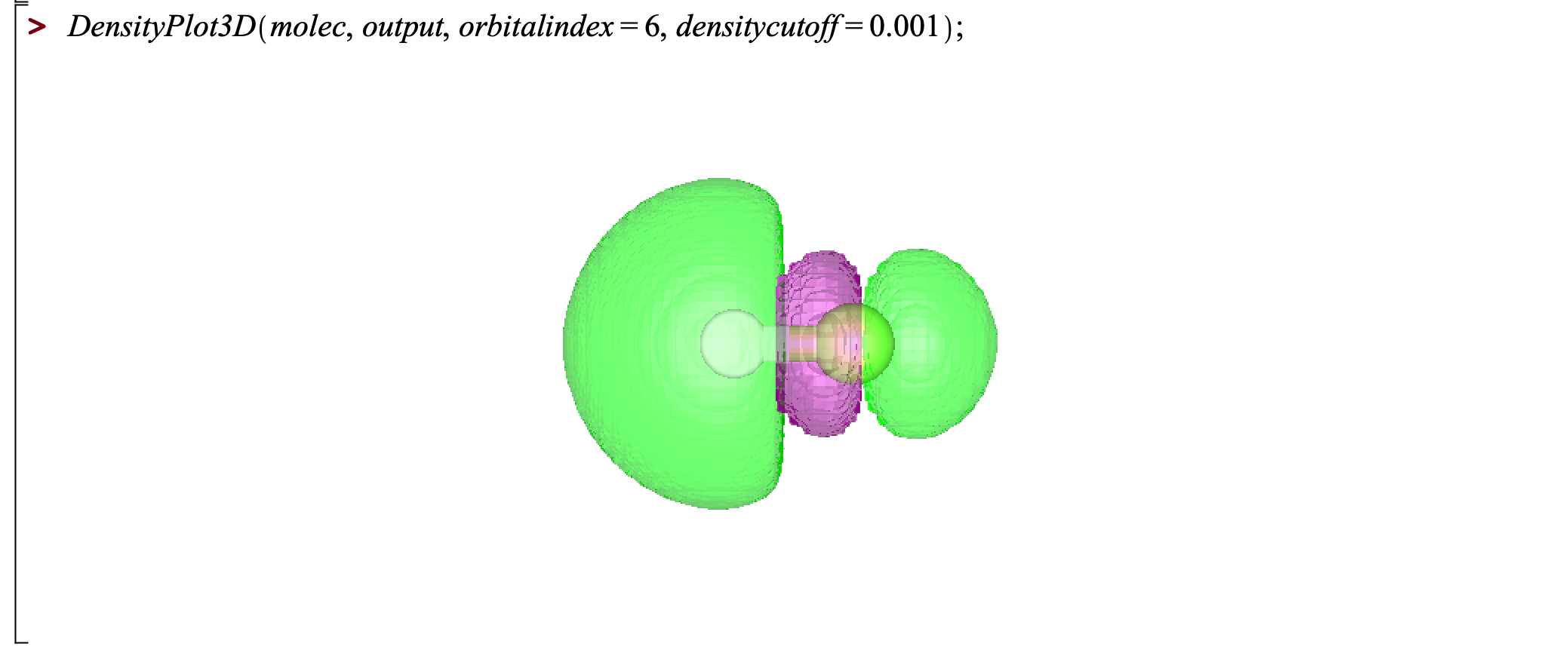} \\

From a pedagogical standpoint, the activity lends itself to simple visualization of an abstract concept, such as MOs, or to the ability for students to explore more complex connections between a model chemistry (electronic structure method and  underlying atomic orbital basis set) and the resulting energies and molecular properties.


\subsection{A Particle in a Symmetric Box}

\label{sec:particle-box}

For this activity, we use Maple's built-in plotting functions to visualize the energies and wavefunctions of a symmetric particle-in-a-box (PB). We then apply the PB model to two chemistry applications, one a toy model and one that corresponds to the ubiquitous conjugated dye lab.\\

\noindent \ \includegraphics[scale=0.50]{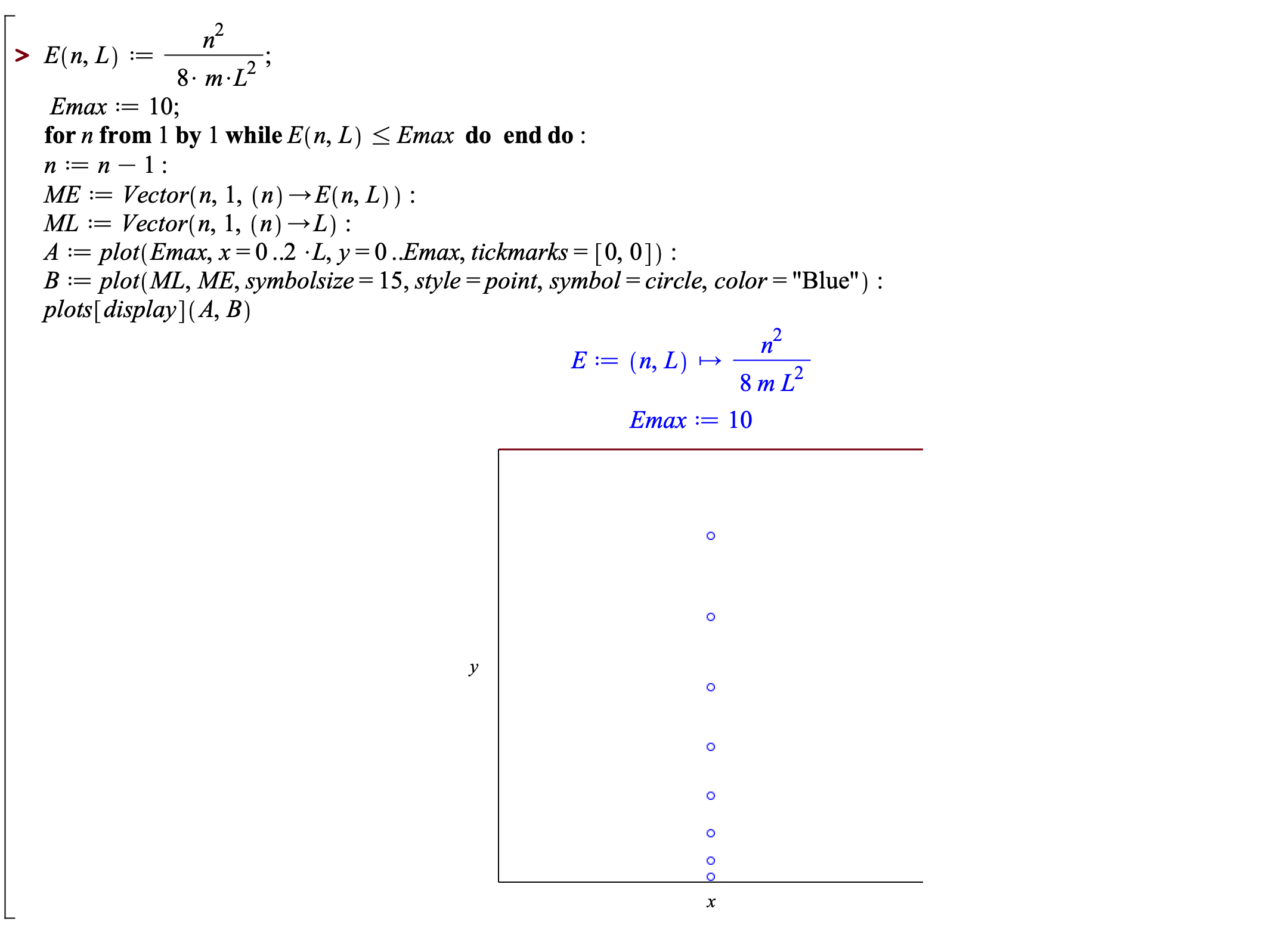} \\

\noindent \includegraphics[scale=0.45]{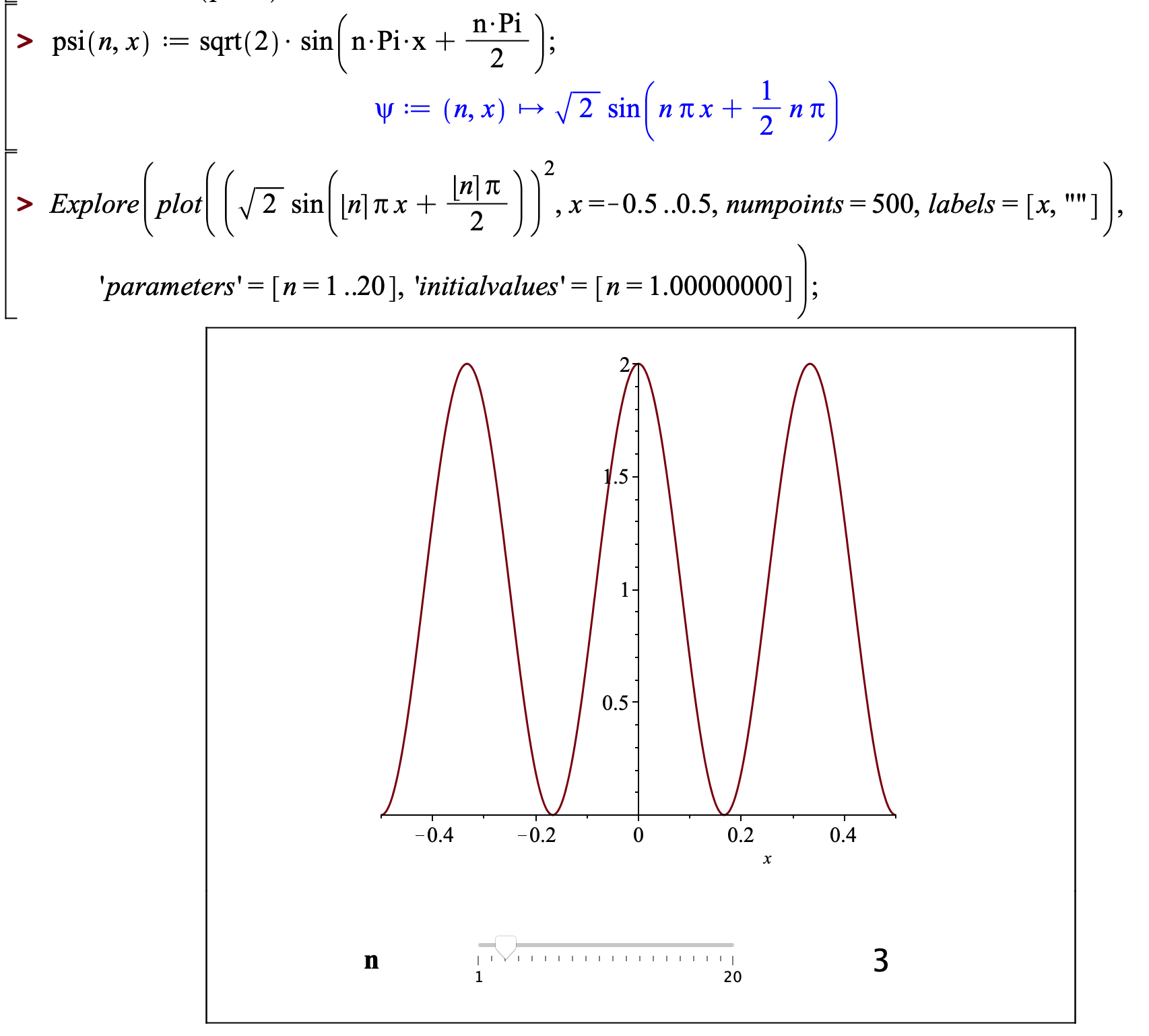} \\

The advantage of having students use Maple's plotting functions to visualize PB energies and wavefunctions is to provide the necessary background and understanding of the relationship between higher energy states and their corresponding nodal structure.  To reinforce the notion that the PB model can be useful in understanding real chemical systems, we use the QCP to calculate the energies and wavefunctions of a chain of H-atoms.  As seen in Sec.~\ref{sec:HFMOs},  the user can input the geometry using a list of lists (output suppressed) and visualize the molecule using the {\em PlotMolecule} function.\\

\noindent \ \includegraphics[scale=0.50]{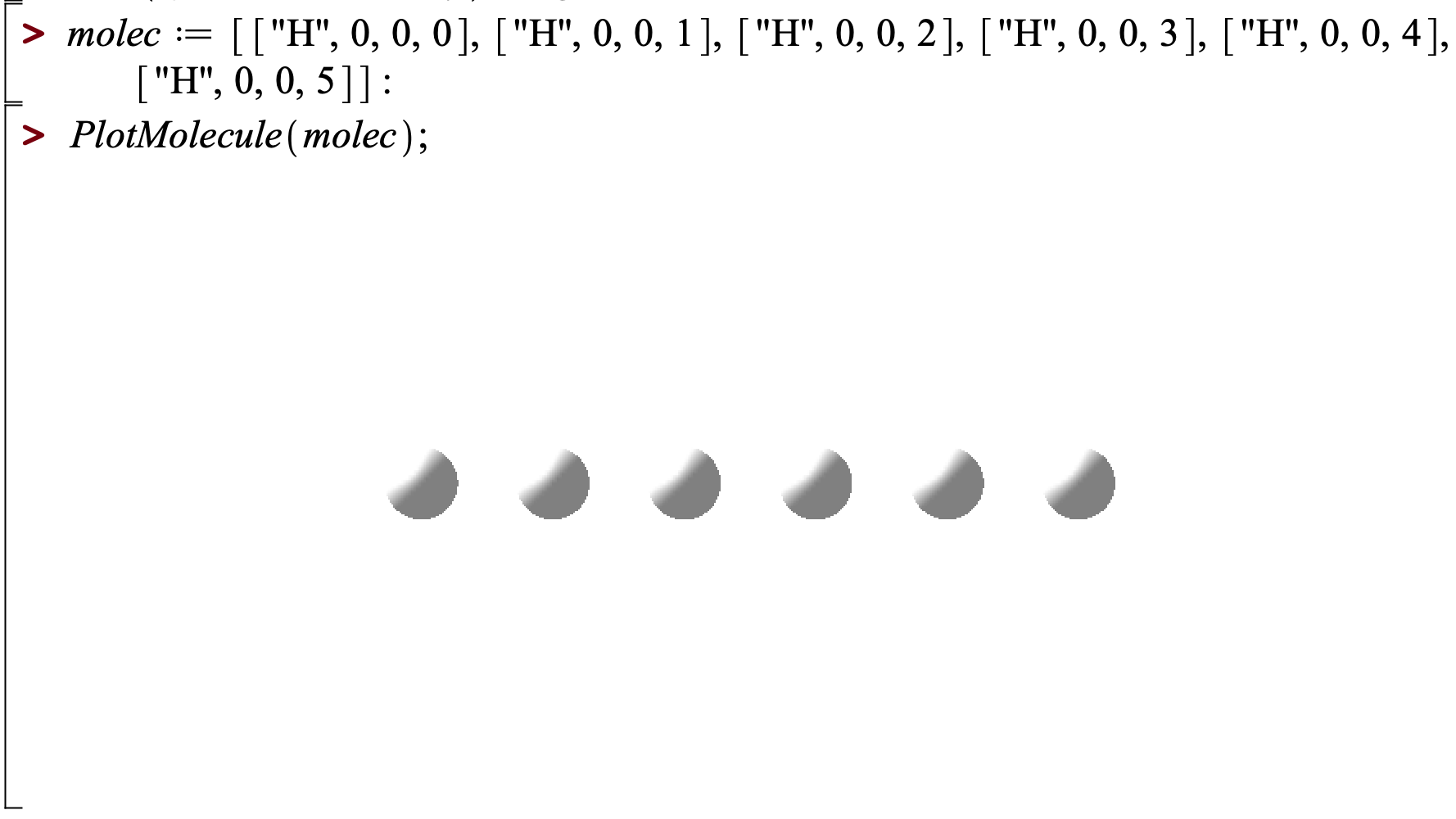}

\noindent In this activity, we use the Hartree-Fock method with a minimal STO-3G basis to calculate the energy and molecular orbitals. The output here has been suppressed, but an example for the HF molecule is provided in Sec.~\ref{sec:HFMOs}.  \\

\noindent \ \includegraphics[scale=0.5]{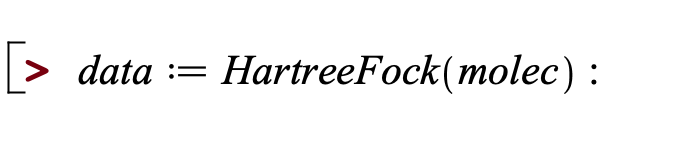}

\noindent The user can visualize molecular orbitals using the {\em DensityPlot3D} command and compare the nodal structures to those expected based on the simple PB model. \\

\noindent \ \includegraphics[scale=0.50]{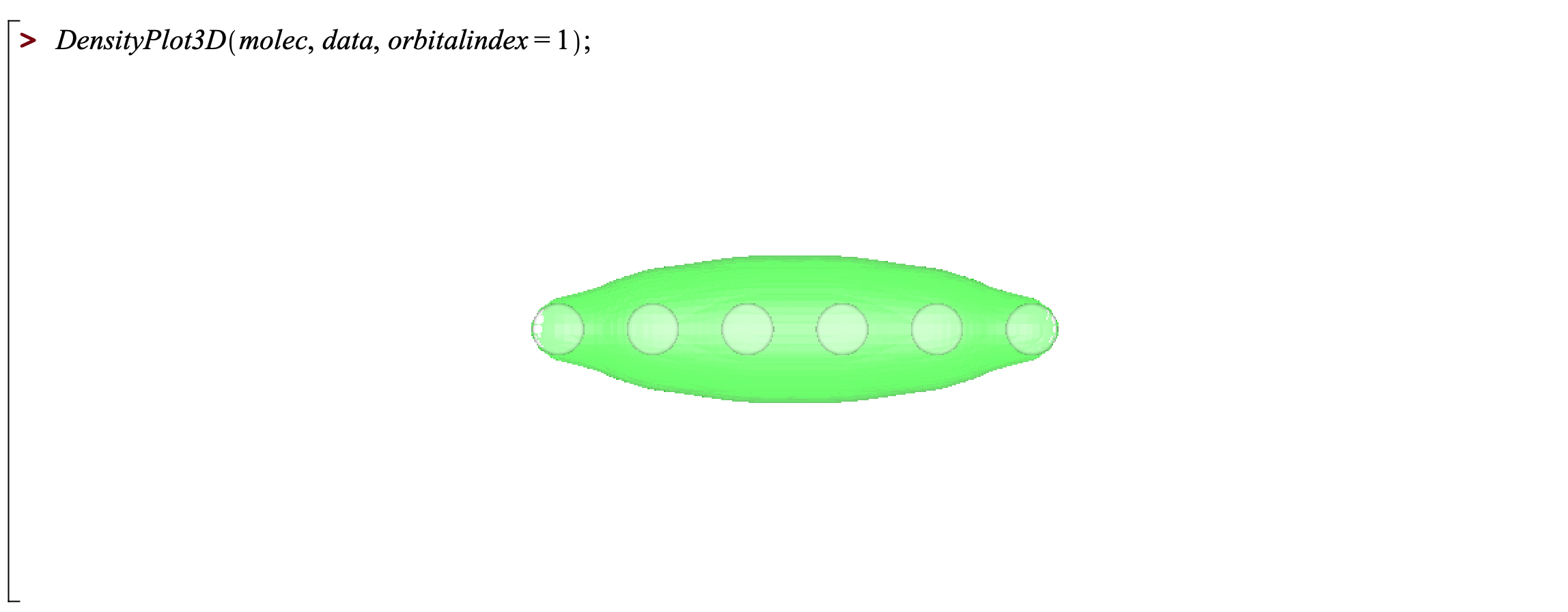}

\noindent \ \includegraphics[scale=0.50]{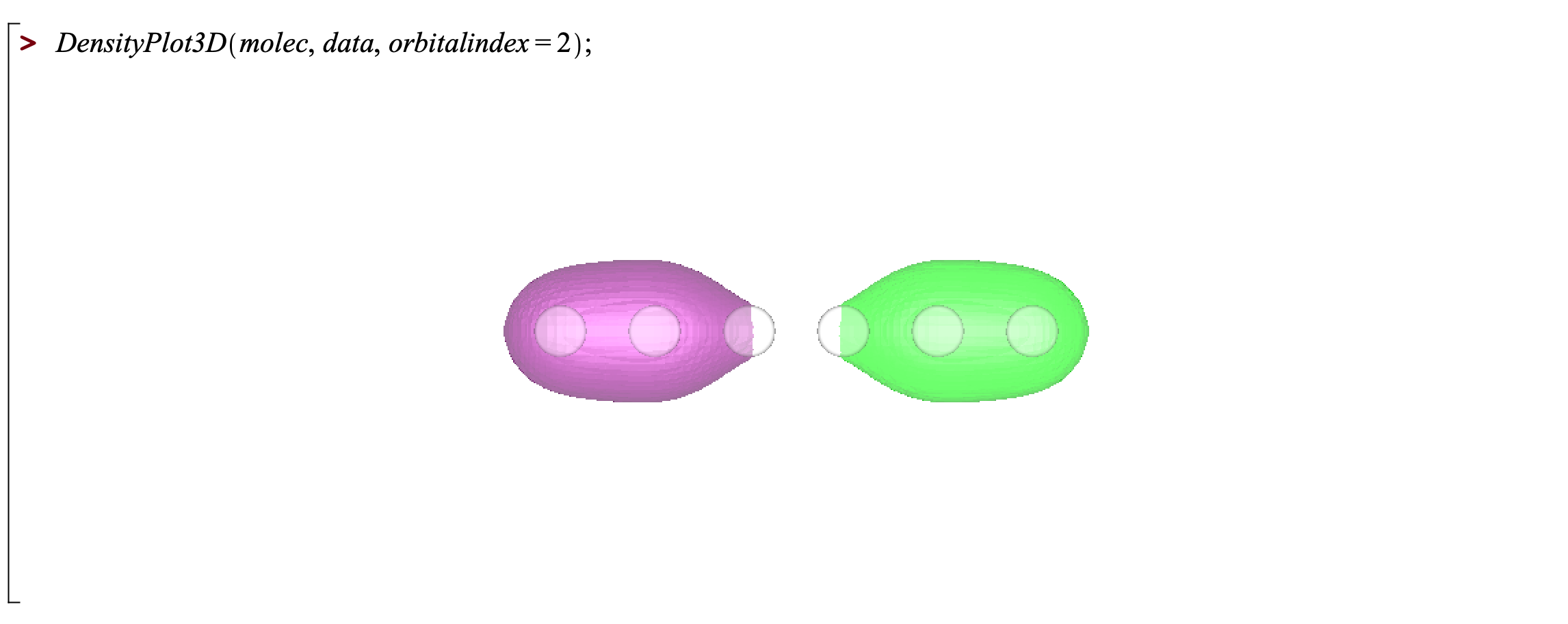}

\noindent \ \includegraphics[scale=0.50]{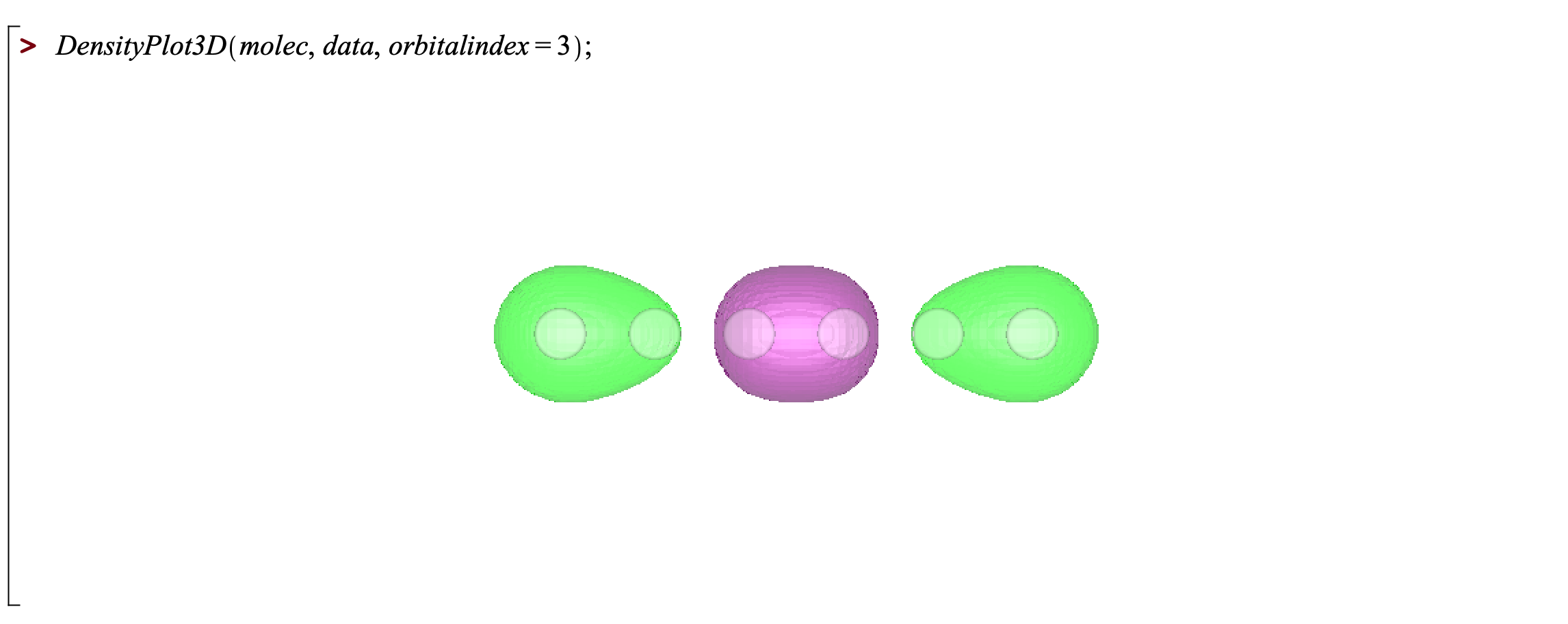}

\noindent \ \includegraphics[scale=0.50]{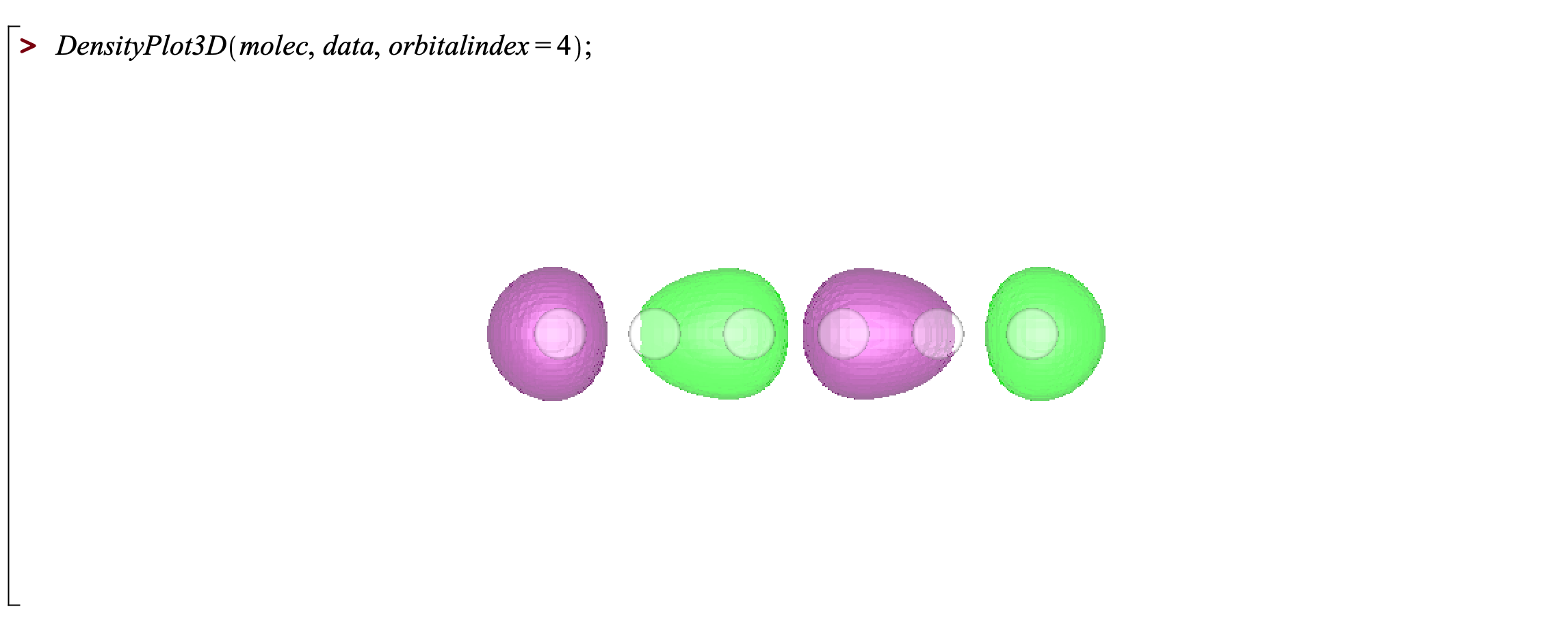} \\

In another application of the PB model,  we again use the Hartree-Fock method and a minimal basis to calculate the energies and MOs for a series of cyanine dyes, the structures of which are provided. Here we show the 3,3'-diethylthiacarbocyanine iodide dye and the resulting HOMO. \\

\noindent \ \includegraphics[scale=0.50]{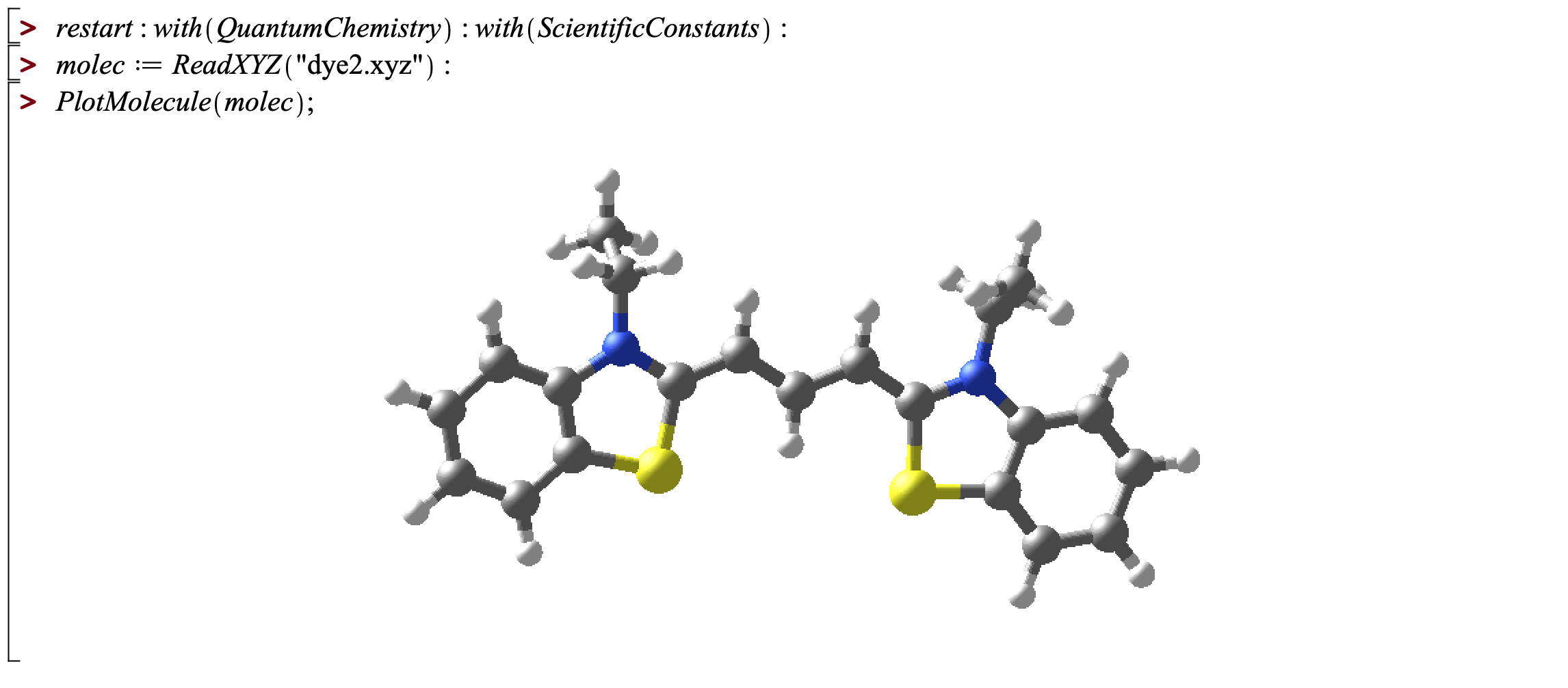}

\noindent \ \includegraphics[scale=0.45]{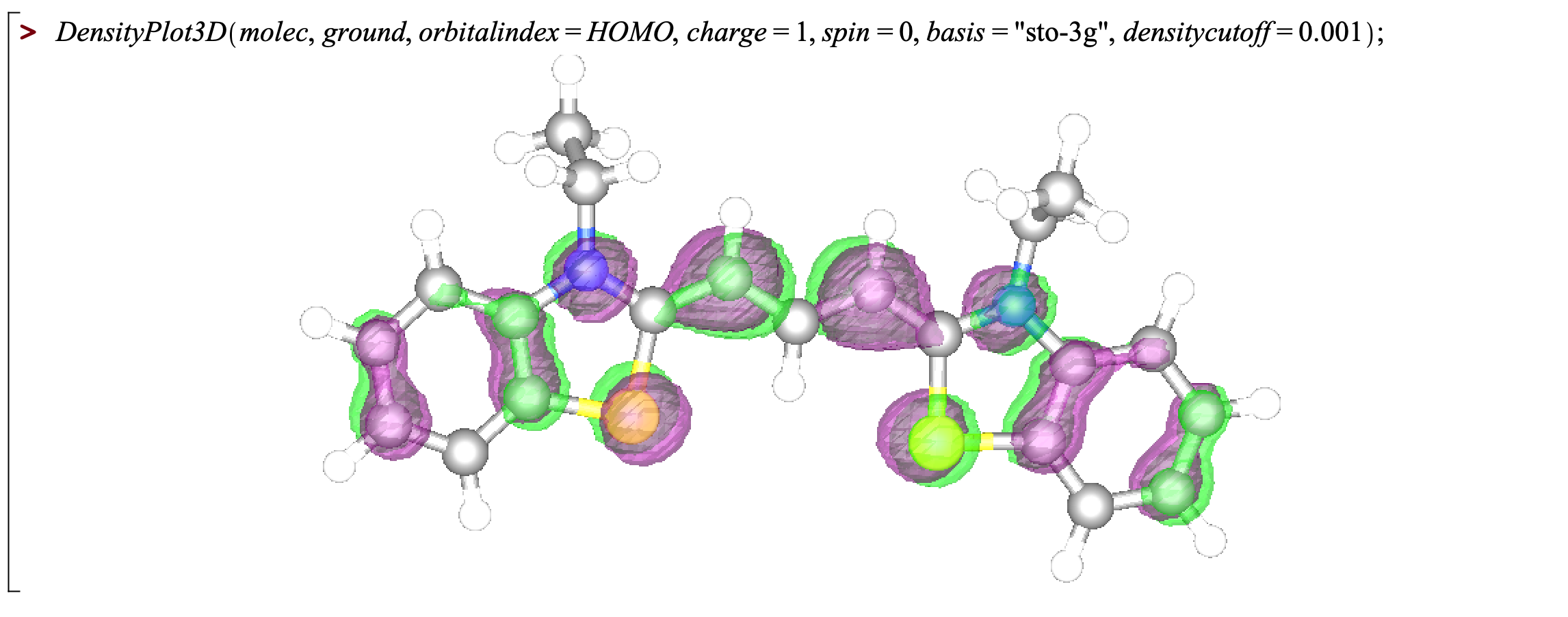} \\

\noindent  The nodal structure along the conjugated chain can again be compared to the $n = 4$  PB wavefunction, which would correspond to the HOMO for a PB system with 8 $\pi$-electrons.  In the activity, students can also {\em approximate} peak absorption wavelength by approximating $\Delta E = E_\mathrm{LUMO} - E_\mathrm{HOMO}$.  Upon completion, students can see that 1) peak wavelengths calculated using electronic structure methods are in qualitative agreement with experiment, and 2) higher level electronic structure models, such as density functional theory (DFT)\cite{parr,kohn,jones} and 6-31G basis set, can improve accuracy.


\subsection{Geometry Optimizations and Normal Modes}

\label{sec:normal-modes}

Another important concept for students, from general to organic to physical chemistry, is the notion of an optimum molecular geometry and associated vibrational modes.  In this activity, students calculate the optimum geometry and normal modes of hypochlorous acid (HClO).  Students can specify a starting geometry in bond-angle coordinates (for example $r_\mathrm{HO}=1.0$, $r_\mathrm{OCl}=1.7$, $\theta\mathrm{(HOCl)}=104.5$), an electronic structure method (default = Hartree Fock) and atomic orbital basis set (default = STO-3G).   Upon completion, students can compare structures before and after optimization using the {\em BondDistances} and {\em BondAngles} functions.\\

\noindent \ \includegraphics[scale=0.5]{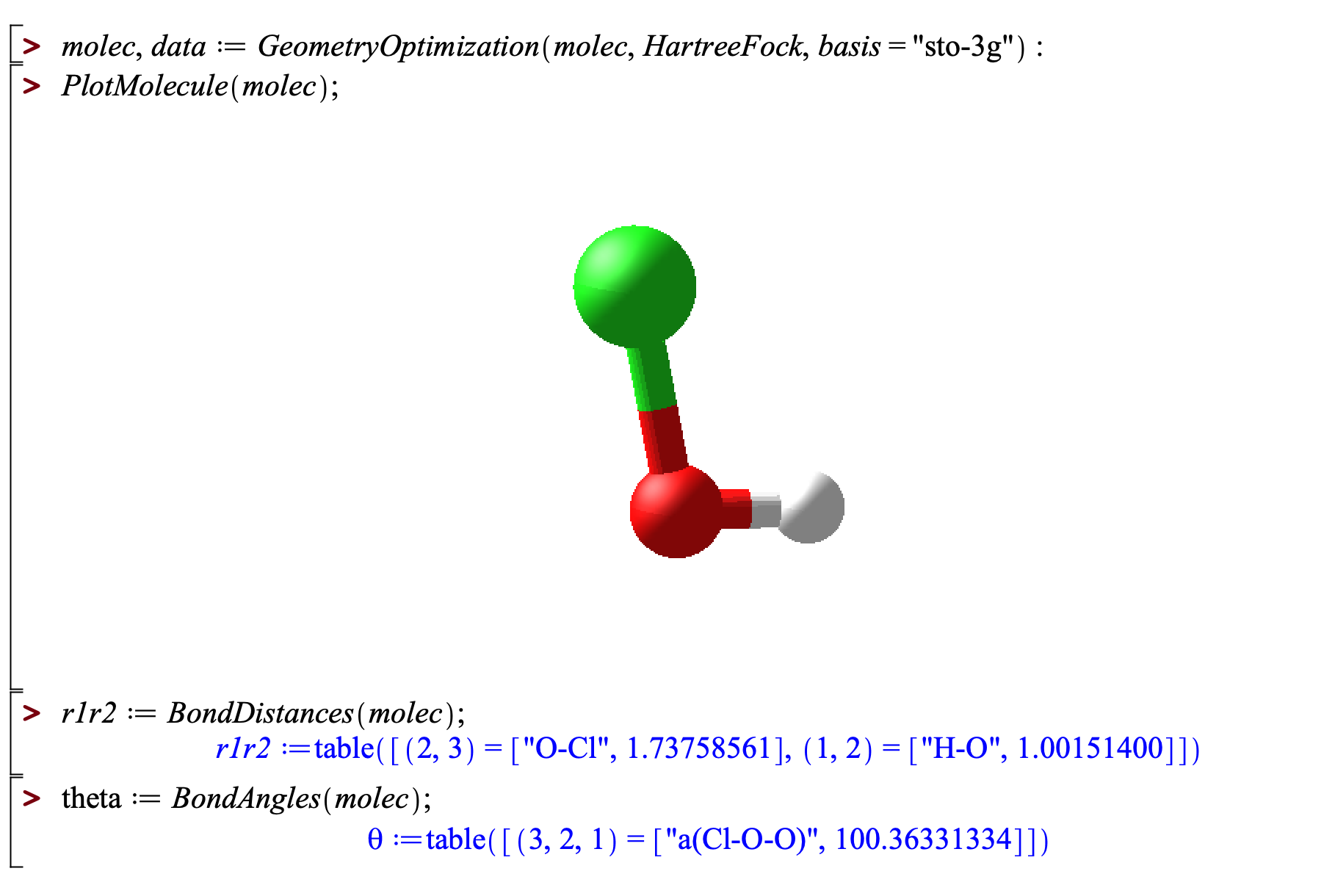} \\

A follow-up calculation for the normal modes is accomplished using the {\em VibrationalModes} command. Students can animate each vibration using the {\em VibrationalModeAnimation} command. (Here, we have modified output to show three frames of the animation for printing purposes.) \\

\noindent \ \includegraphics[scale=0.45]{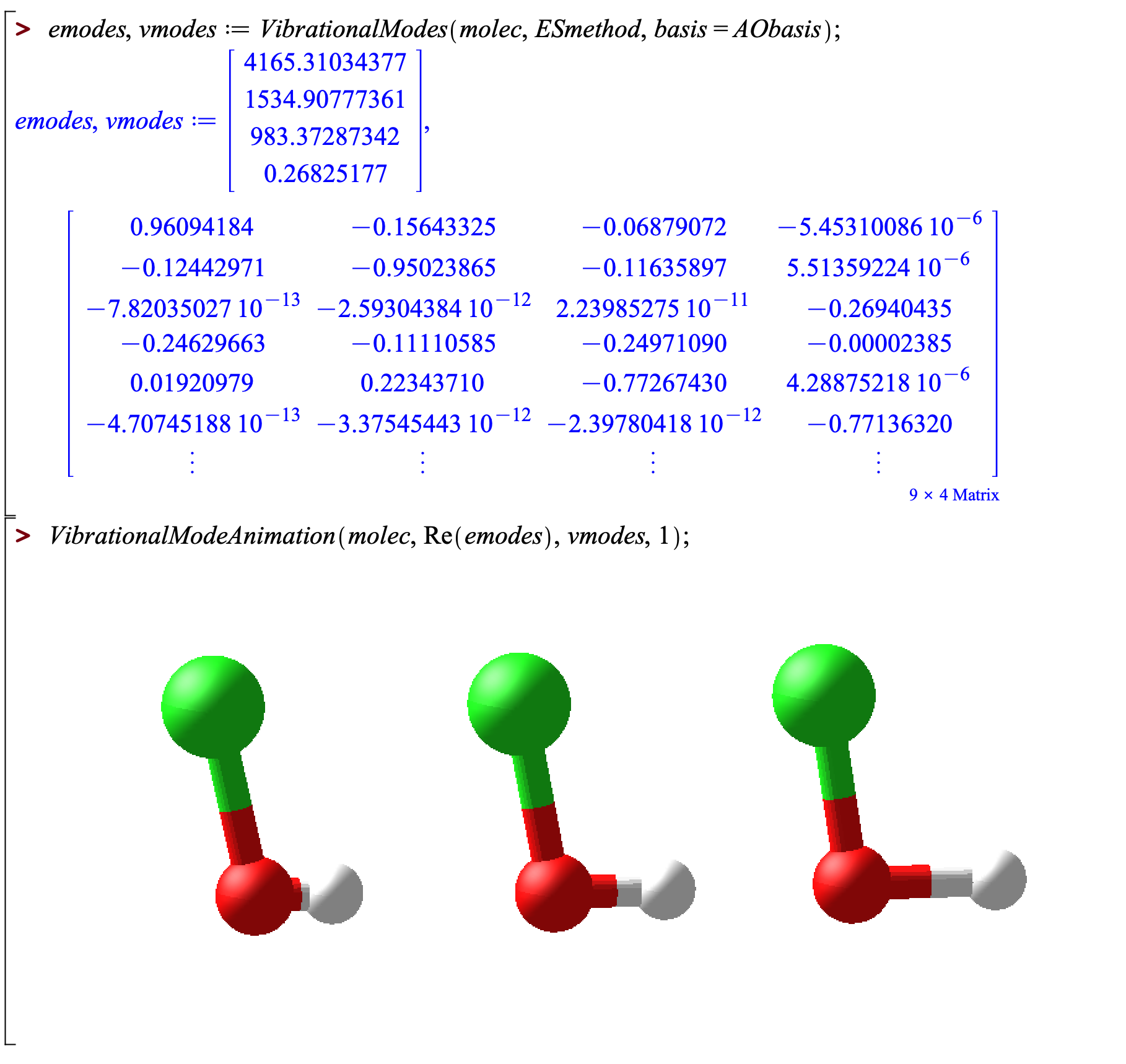} \\

\noindent Students can compare calculated frequencies with literature values (3609 cm$^{-1}$ (OH stretch), 1239 cm$^{-1}$ (bend), and 724 cm$^{-1}$ (Cl-O stretch)) and can explore higher level model chemistries to get better agreement with experiment.


\subsection{Calculating Reaction Thermodynamics for Combustion of Methane}

\label{sec:thermo}

It is important to note that the utility of the QC package extends beyond only explicit electronic structure applications.  For example, in one activity provided with the package, we use the {\em Thermodynamics} function to calculate the thermodynamics of combustion for methane:
\begin{equation}
\mathrm{C}\mathrm{H}_4\mathrm{(g)} +\mathrm{O}_2\mathrm{(g)} \longrightarrow \mathrm{C}\mathrm{O}_2\mathrm{(g)} + \mathrm{H}_2\mathrm{O}\mathrm{(g)}\ \ \ \ \ \ \ \Delta H_\mathrm{rxn}=?, \Delta G_\mathrm{rxn}=?, \Delta S_\mathrm{rxn}=? \label{rxn1}
\end{equation}
For a given reaction specie in Rxn~\ref{rxn1}, for example CH$_4$, the user specifies the name and corresponding label, symmetry number, electronic structure method and basis with which to calculate energies, electronic structure and basis with which to calculate the vibrational frequencies (usually the same as energy but not required), a scaling factor, and molecular spin corresponding to singlet, doublet, triplet, etc.: \\ \\
\noindent \ \includegraphics[scale=0.6]{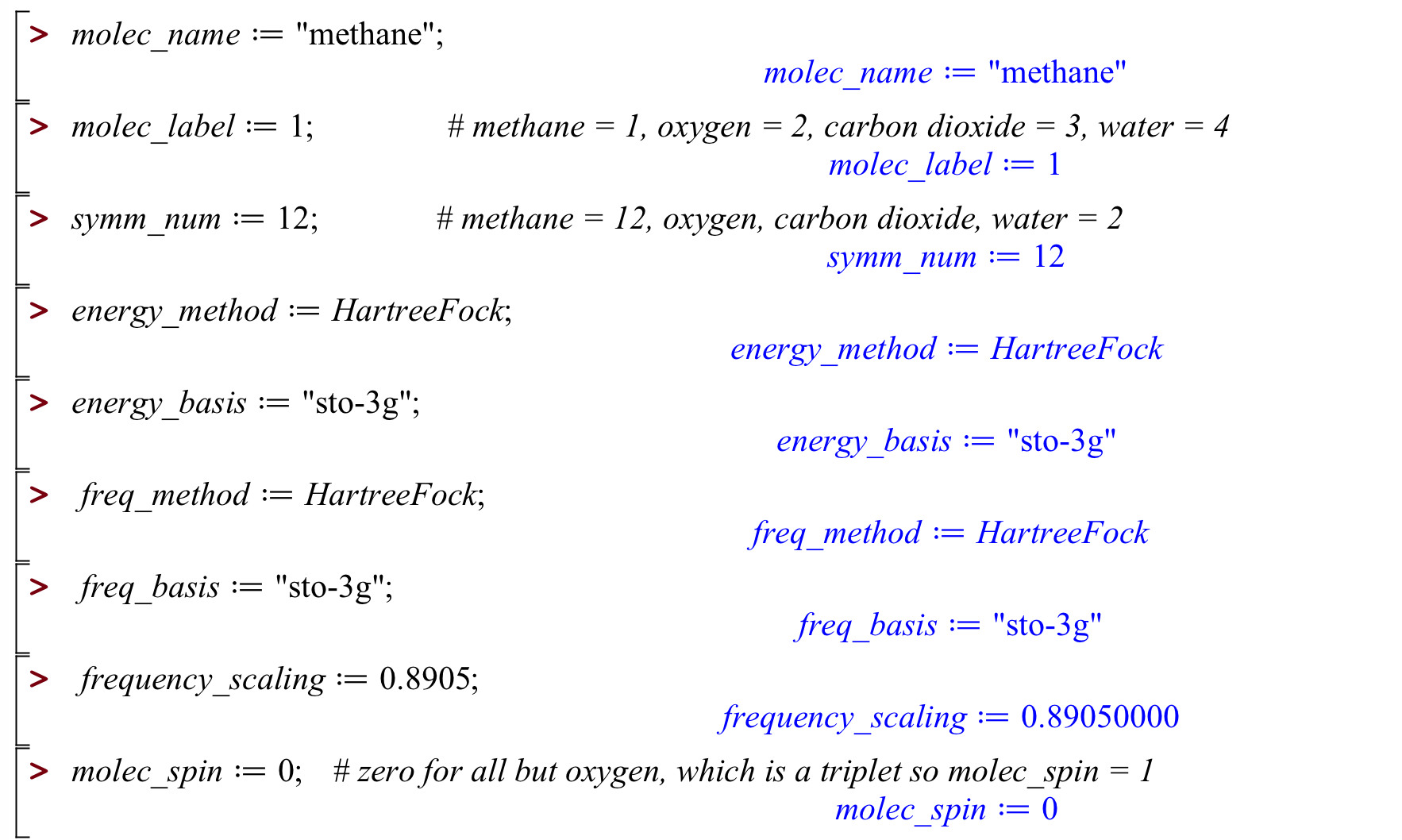} \\
Once a given molecular specie has been specified, the user can load the molecule using the {\em MolecularData} function, which retrieves a 3D geometry from PubChem based on the name or CID provided and execute the {\em Thermodynamics} function to calculate the thermodynamic variables $S$, $H$, and $G$:\\ \\
\indent \includegraphics[scale=0.5]{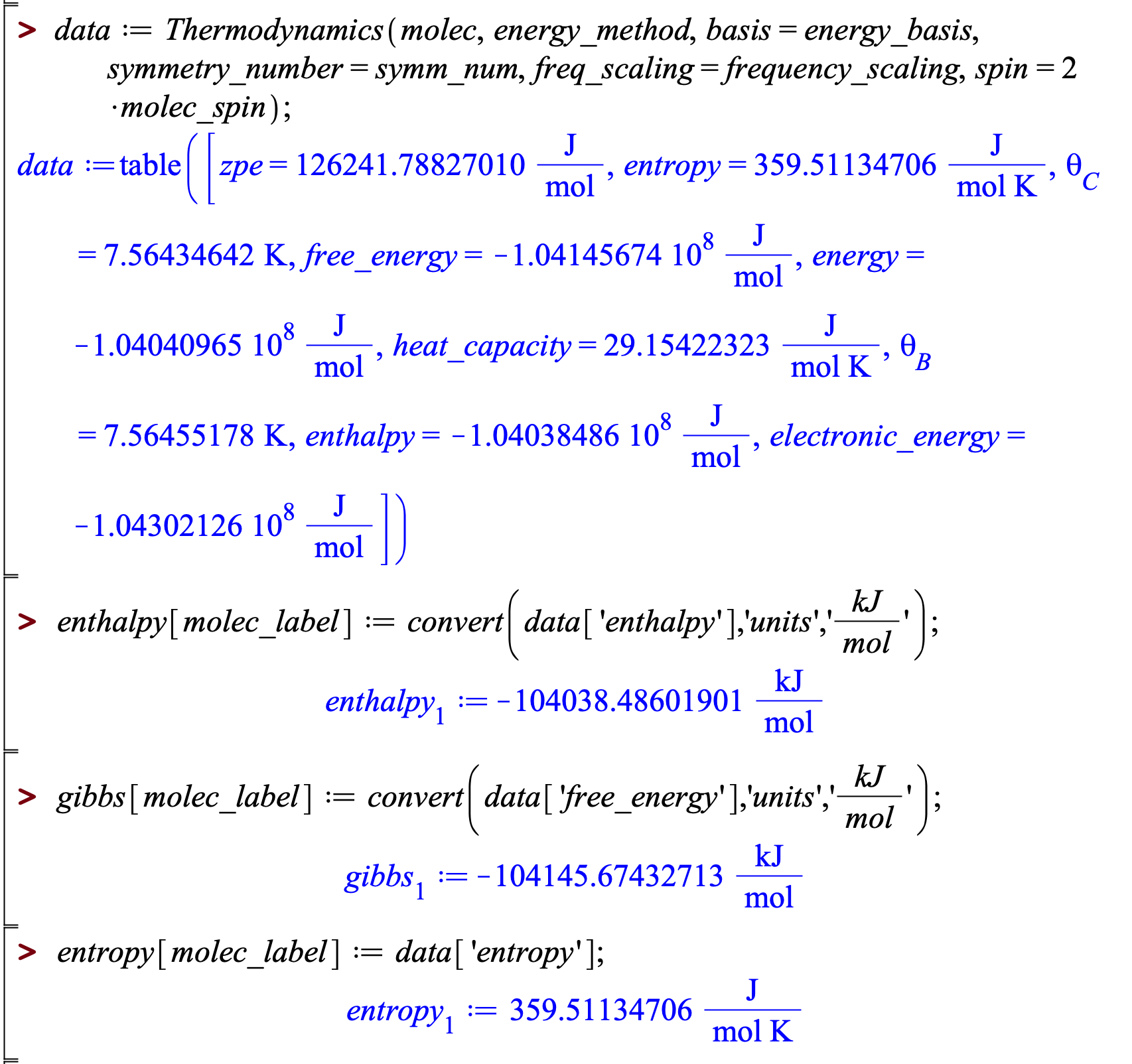} \\

The above input and subsequent calculation output would be updated for each of the molecules in Rxn~\ref{rxn1}. Once thermodynamic variables for each of the reactant and product species is calculated, the entropy of combustion ($\Delta S$), enthalpy ($\Delta H$), and Gibbs free energy ($\Delta G$) can be calculated using Hess's Law:
\begin{equation}
\Delta F = \sum_i n_iF_{product,i} - \sum_j n_j F_{reactant,j}
\end{equation}
where $F$ is the desired thermodynamic variable, and $n_i$ and $n_j$ are the molar coefficients in Rxn~\ref{rxn1}:\\

\indent \includegraphics[scale=0.55]{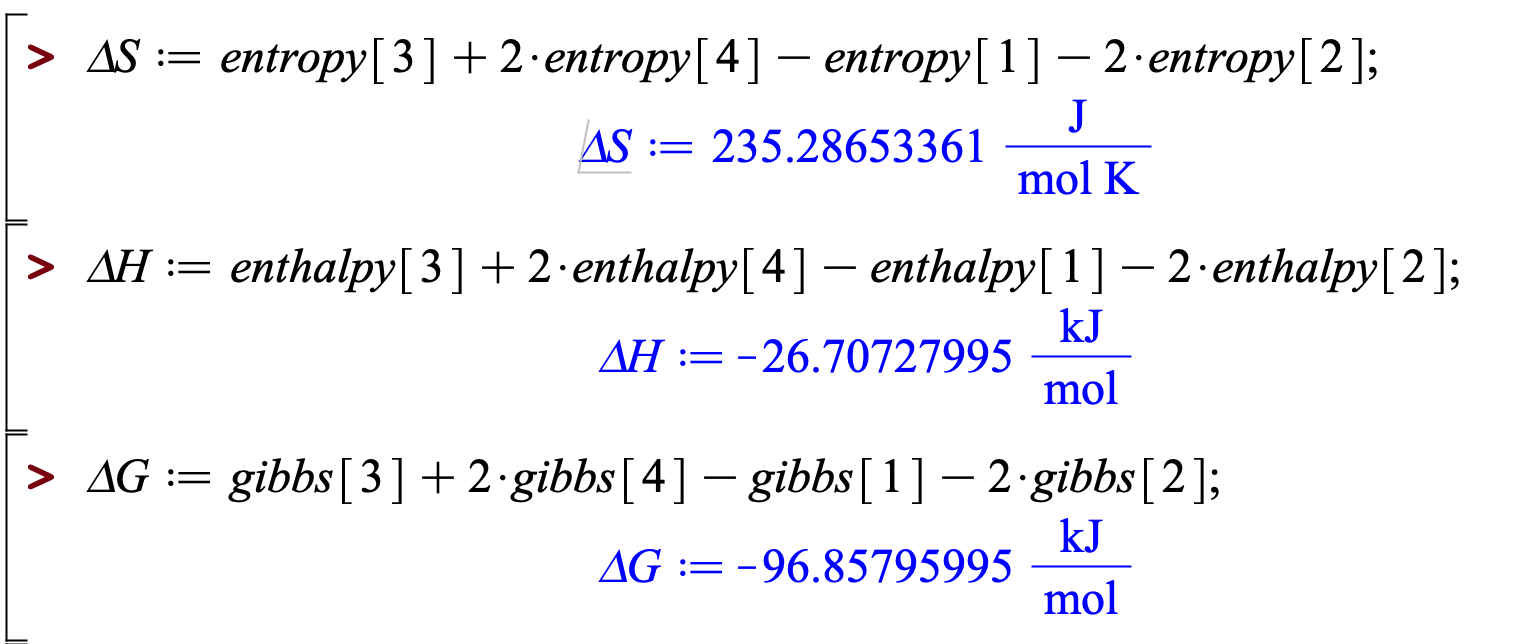} \\

These results can be compared with predictions using Hess's law and tabulated standard entropy and standard enthalpy and free energy of formation values.

\section{Discussion and Conclusions}

\label{sec:discussion}

In this paper, we have introduced the Quantum Chemistry Package, implemented in Maple, and have shown how it can be used in the chemistry classroom.  It is important to emphasize that the QCP contains both wavefunction and density functional methods.  Electronic structure methods highlighted above include the Hartree-Fock and DFT methods, but a diverse set of electron correlation methods are also included in the QCP, such as  2nd-order M{\o}ller-Plesset perturbation theory (MP2)\cite{moller}, coupled cluster\cite{purvis,bartlett}, full and active space variants of configuration interaction (CI).\cite{knowles,olsen,helgaker,roos}  Furthermore, unique to the package are two-electron reduced density matrix (2RDM) methods, parametric 2RDM (P2RDM)\cite{mazziotti2,mazziotti3,foley,valentine} and variational 2RDM (V2RDM),\cite{mazziotti4,mazziotti5,schlimgen,montgomery} which are well-suited for strongly correlated molecules where they can accurately describe quantum effects that are difficult to treat by conventional methods.  The breadth of methods available in the package also make it suitable for research grade calculations.

It is our hope that instructors from all levels of chemistry use the imbedded lessons as well as create their own to share with the chemistry and Maple communities in order to help integrate quantum calculations throughout the chemistry curriculum!

\begin{acknowledgement}

D.A.M. gratefully acknowledges  the United States Army Research Office (ARO) Grants W911NF-16-C-0030 and W911NF-16-1-0152 and the United States National Science Foundation Grant CHE-1565638.

\end{acknowledgement}




\bibliography{refs}

\end{document}